# Current Assisted Magnetization Switching in (Ga,Mn)As Nanodevices

C. Gould, K. Pappert, C. Rüster, R. Giraud*, T. Borzenko, G. M. Schott, K. Brunner, G. Schmidt and L.W. Molenkamp

Physikalisches Institut (EP3), Universität Würzburg, Am Hubland, 97074, Würzburg, Germany

Current induced magnetization switching and resistance associated with domain walls pinned in nanoconstrictions have both been previously reported in (Ga,Mn)As based devices, but using very dissimilar experimental schemes and device geometries . Here we report on the simultaneous observation of both effects in a single nanodevice, which constitutes a significant step forward towards the eventual realization of spintronic devices which make use of domain walls to store, transport, and manipulate information.

As the field of semiconductor spintronics [1] continues to mature and more and more of the fundamental issues become resolved, researchers are increasingly turning their attention to harnessing the spin transport properties of magnetic semiconductors into industrially relevant devices. One widely explored avenue towards the design of spintronic devices with novel functionalities is based on using the position of individual domains in ferromagnetic semiconductors (FS) to store information, and the controlled movement of domain walls to manipulate and transport this information. For such devices to be industrially relevant, schemes are needed that are scaleable to large integration density, where the position of domain walls can be determined by electrical measurements, and where the walls can be manipulated using gate or bias voltages. One paradigm for such a device is the "race-track memory" concept which was disclosed by IBM in February 2005 [2].

* Present address: Laboratoire de photonique et de nanostructures, CNRS, 91460 Marcoussis, France

Current induced switching is well established in metals [3,4], but currently operates only at switching current densities of $10^7$-$10^8$ A.cm$^{-2}$, exceeding the value tolerated in modern integrated circuit technology [5,6]. It has however been predicted [7,8] theoretically, and recently demonstrated experimentally [9], that current induced switching in FS can be achieved for current densities 2 to 3 orders of magnitude lower than in metals. In that experiment, a variation in layer thickness was used to demark the various regions of a (Ga,Mn)As device, and current induced switching of a central region was observed by magneto-optical Kerr effect (MOKE) and anomalous Hall measurements in a device operating a few degrees below its Curie temperature using a current density of $8x10^4$A.cm$^{-2}$.

A second potentially important advantage of FS over their metallic counterparts was the prediction [10,11] that domain wall resistance in FS should be significantly larger than in metals. This has also been experimentally confirmed in (Ga,Mn)As devices [12], where a large change in device resistance was observed, associated with the pinning of a domain wall in a nanoconstriction between regions of oppositely magnetized (Ga,Mn)As.

In the present work, we show results on samples which combine both of the above elements in a single device and show that they can be mutually compatible. Our device is a lithographically defined nanostructure patterned in a 20 nm thick low temperature $Ga_{0.94}Mn_{0.06}As$ layer grown on a GaAs buffer. The (Ga,Mn)As layer is thus compressively strained, and has magnetic easy axes in the plane of the layer along the [100] and [-100] crystal directions. The layer has an as-grown Curie temperature of ~70 K as determined by SQUID (superconducting quantum interference device). This layer is then patterned into nanostructures as shown in the SEM image and schematic diagram inset in Fig. 1. The structure is defined using a negative e-beam lithography process, and $Cl_2$ based dry etching. It is comprised of a small central island, which is separated from large triangular leads by a pair of nanoconstrictions of ~10 nm in width, and is oriented such that the current path is along an easy axis.

Figure 1 shows a 4.2 K magnetoresistance measurement of the device as a magnetic field applied in a direction parallel to the current path is scanned from -50 to +50 mT and back again with a bias voltage of 5 mV. The sample shows a clear spin valve like signature, which can be explained, as discussed in detail in ref. 12, by the fact that the large leads will have a much lower coercive field than the small island. As such, when the magnetic field is swept, there exists a field range for which the magnetization in the leads has already reversed while that of the island remains fixed. In this configuration, the domain walls will be pinned in each of the nanoconstrictions, and an increase in resistance associated with these domain walls is observed.

We now present evidence that, after setting up the device in this configuration where the central region is anti-parallel to the leads, we can, without further changing the magnetic field, reverse the polarization of the island using a current assisted switching mechanism. This is shown in Fig. 2, where we plot four magnetoresistance measurements. The first, indicated by the solid line is a normal magnetoresistance measurement as in Fig. 1, again taken with a bias voltage of 5 mV. The second scan, with the data displayed as dots, is a similar measurement, except that the scan is stopped at 8 mT (indicated by the vertical dotted line), and several minutes are allowed to elapse before the scan is continued. This interruption has no effect on the measurement. The final two curves are however more interesting. Again, we repeat a similar magnetoresistance measurement, which is stopped at 8 mT. However, this time, instead of simply waiting, we ramp the bias voltage up to +100 mV (dark thick line) or -100 mV (light thick line), and back down to 5 mV before continuing the sweep. The time taken for this process is shorter than the waiting time used for the second curve, but now, when the curves are continued, it is apparent that the central region has already reversed its magnetization as the device is in the low magnetoresistance state. Note that the sign of the magnetization reversing voltage is unimportant as the device and magnetic configuration are symmetric.

While these measurements are done at 4.2 K [13], well below the Curie temperature of the material, and the total power applied during the current induced reversal is only some 100 nW, because of the small size of the region being switched, one might speculate that local heating is playing a role, warming the (Ga,Mn)As locally above its Curie temperature, and then having it magnetize in the opposite direction as it cools upon removal of the high bias. This hypothesis can however be ruled out by the data presented in Fig. 3. Here we first prepare the sample in the anti-parallel configuration as in Fig. 2, by starting a magnetoresistance scan, and interrupting it at 8 mT. After interrupting the magnetoresistance scan, we remain at this fixed field, and perform a series of resistance versus voltage measurements. We first measure the resistance while increasing the bias up to 180 mV (dark open circles), and notice, in addition to the smooth and monotonic decrease expected from the non-linear nature of the nanoconstriction resistance, some sharp switching events in the region ~100 mV. Then we measure the resistance as we sweep the bias back (dark solid curve), and find only a smooth increase, without any jumps.  At low bias, the two curves differ in resistance by the amplitude of the spin-valve signal. If we now repeat similar bias scans, we obtain the solid curve, both for the sweep up and sweep down, since the sample is in the parallel state, and the high bias voltage can have no further effect. While it is difficult to get exact numbers for the dimensions of the constrictions, using reasonable estimates, we get a typical switching current density of $\sim 1 \times 10^6$ $A.cm^{-2}$.

Repeating the full measurement several times by preparing the anti-parallel state before each sweep up in bias does yield qualitatively reproducible results, in the sense that sharp switching events are observed in every attempt. A second such measurement is shown as the light data in Fig. 3 (The curves having been offset for clarity). Note that the positions of the switching events vary from one data set to the next (and occasionally, even the number of switching event varies). This is probably due to the fact that the nanoconstrictions, due to their small size, are extremely sensitive to minute imperfections in their side-walls. As a result, each nanoconstriction is seeded with multiple, nearly equivalent, pinning centers, which act to pin the domain wall at slightly different positions in each successive measurement, leading to a different geometrical confinement of the domain wall, and thus a different domain wall resistance. Indeed, a similar effect attributed to multiple pinning centers was observed in the magnetoresistance measurements of ref. 12.

The measurements presented in Fig. 3 were performed at a magnetic field of 8 mT. In Fig. 4, we present the results of similar measurements on a second sample, where we repeat the experiment multiple times, stopping at various points on the magnetoresistance curve indicated in the inset of the figure. The curves at 8.47, 9, 9.53, 10.06, and 10.59 mT all show clear switching behaviour. No abrupt switching was observed at magnetic fields below 6.5 mT (as for example in the 6.35 mT curve from the figure) or above 11mT (not shown). Interestingly, for the entire field range between 6.5 and 11 mT where switching events are observed, there is no correlation between the position (or number) of switching events, and the field at which the experiment was performed. The total amplitude of the sum of all switching events on a given curve does of course depend on the magnetic field at which the measurement is performed since the total amplitude must be consistent with the amplitude of the magnetoresistance spin valve signal.

The mechanism driving the current assisted switching in these devices can not be unambiguously determined. While the results of Fig. 3 preclude the idea that we are heating above Tc, since the discontinuous change in resistance at the switching event is inconsistent with the idea of heating above Tc, which is a smooth phase transition in (Ga,Mn)As, we cannot completely rule out heating as playing a role. One could speculate that local heating of the sample to a temperature below its Curie temperature, but sufficiently high to change its coercive field, causes a lowering of the switching field causing the reversal of the central region under the applied magnetic field. We find this explanation unsatisfactory however, both because we would not expect significant heating at the power levels used, and because under this model, for measurements done at lower field, a correspondingly greater reduction of the coercive field would be needed, which should lead to a strong dependence of the switching bias on the magnetic field, which is clearly incompatible with the results of Fig. 4.

For this reason, we believe that the mechanism most likely at work here is the same as invoked in ref. 9; namely a p-d exchange mediated spin angular momentum transfer between the current carrying itinerant holes, and the localized Mn spin in the central region [8].

In summary, we have demonstrated a current induced reversal of the magnetization in a localized region of (Ga,Mn)As, detected by electrical transport measurements of the resistance associated with domain walls trapped in nanoconstrictions. We believe the combination of these two effects in a single device could be an important step forward towards the eventual realization of ferromagnetic semiconductor spintronics memory devices.

Acknowledgements:
The authors wish to acknowledge the financial support of DARPA, the EU, DFG and BMBF. R.G. acknowledges financial support of the Humbolt foundation.

Figure captions:

Fig. 1: Full magnetoresistance scan at 5 mV for magnetic field parallel to the current showing the spin valve signal of the device. The insets show a scanning electron microscope (SEM) image and a schematic of the device.

Fig. 2: Four magnetoresistance sweeps of the device, after saturating at -50 mT. Three of the sweeps were interrupted at 8 mT as described in the text.

Fig. 3: Resistance versus voltage measurements at 8 mT for the sample initially prepared in the anti-parallel configuration. The sweeps to higher bias show clear switching events which are absent on the sweeps towards lower bias. The sets of curves are offset for clarity.

Fig. 4: Resistance versus bias measurements as in Fig. 3, in different magnetic fields as indicated in the figure. The sets of curves are offset for clarity.

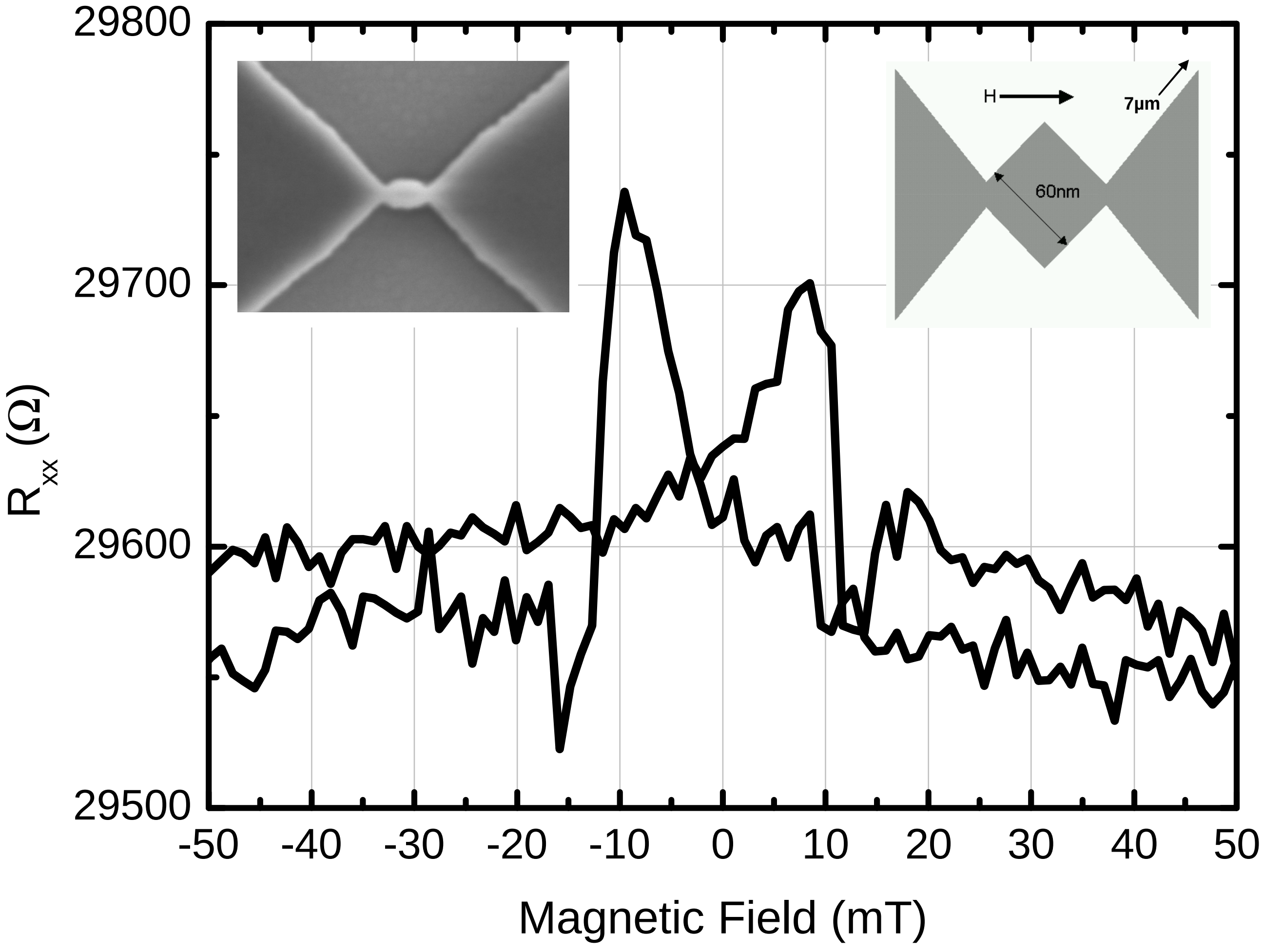

**Fig. 1**

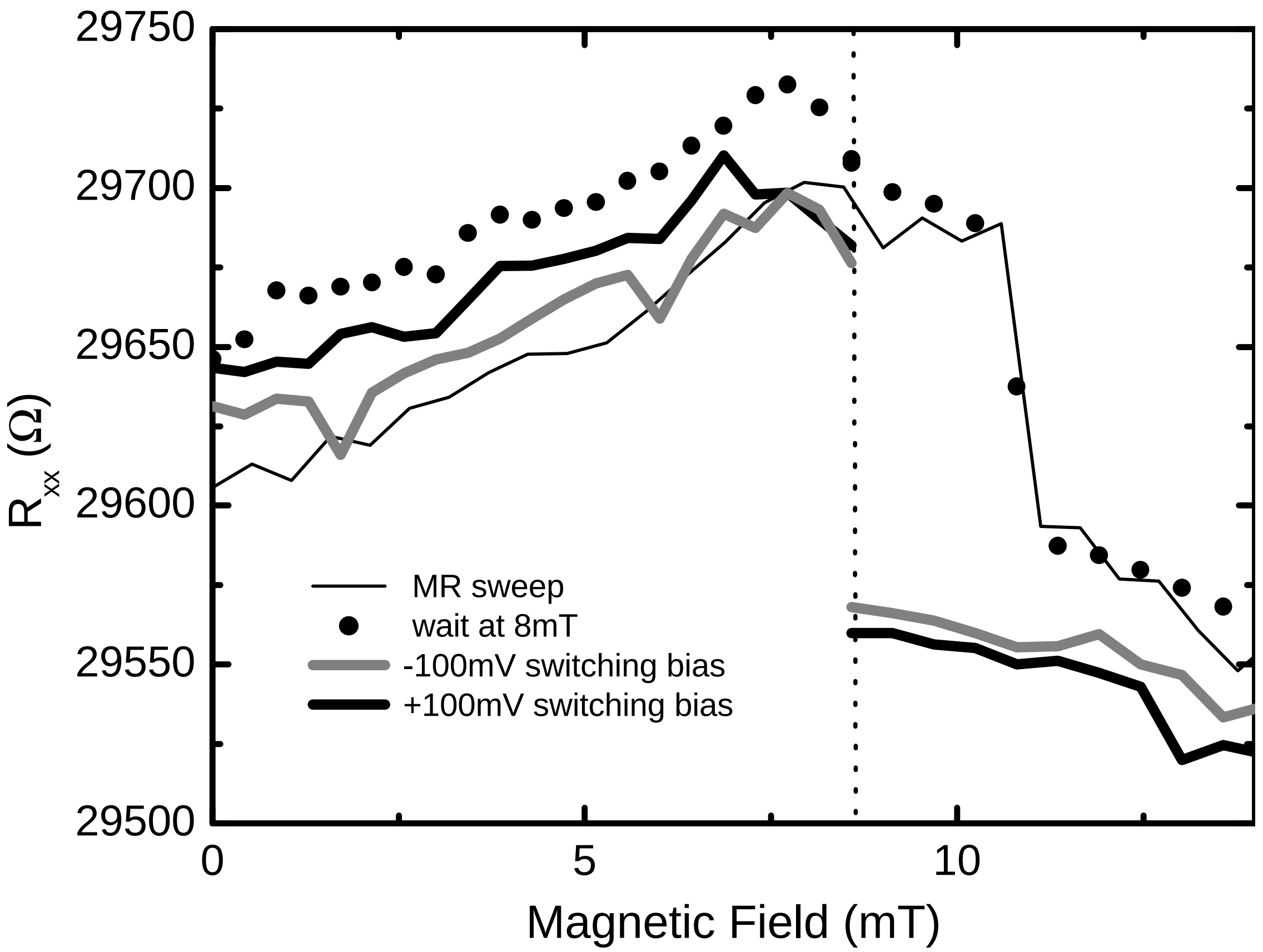

**Fig. 2**

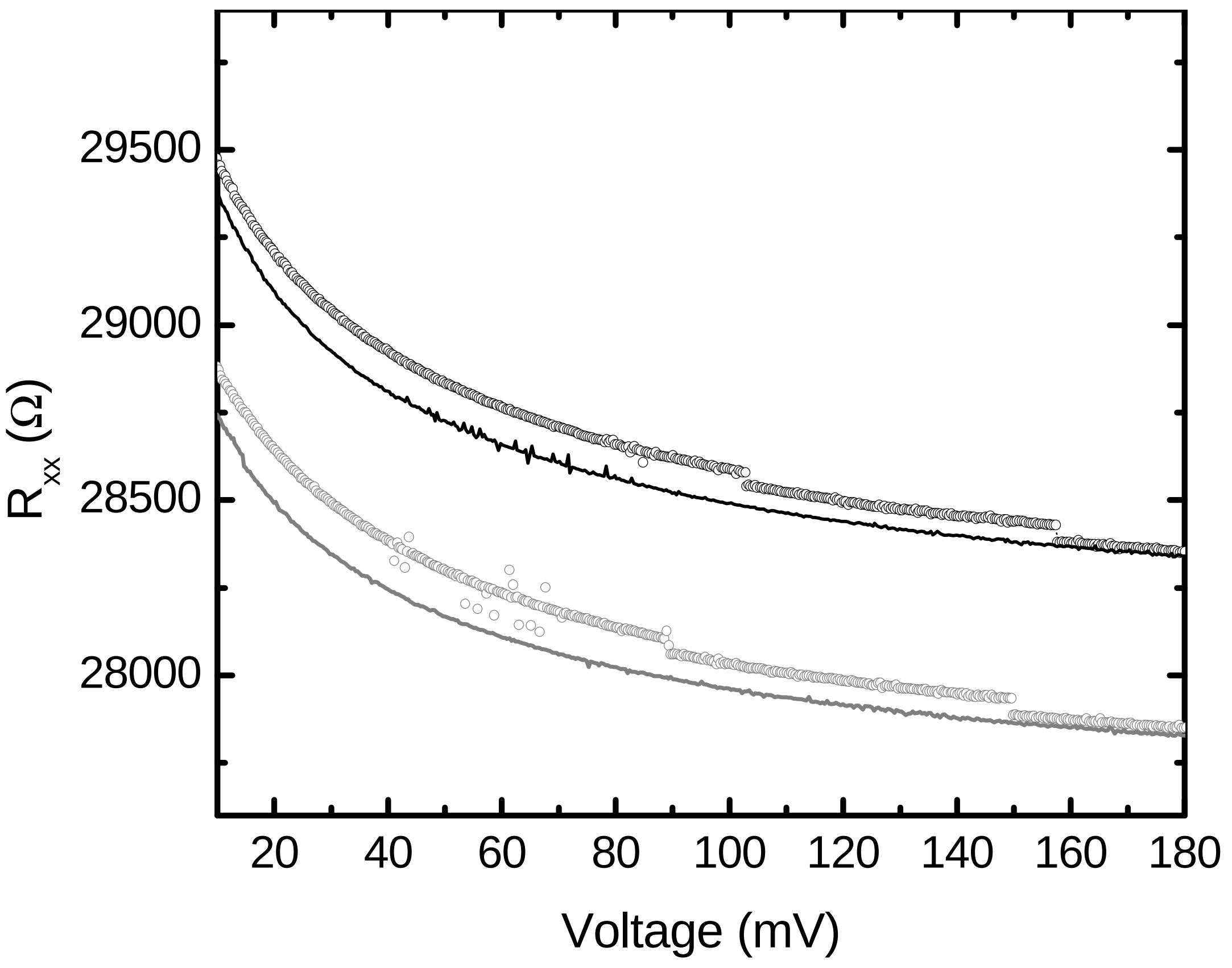

**Fig. 3**

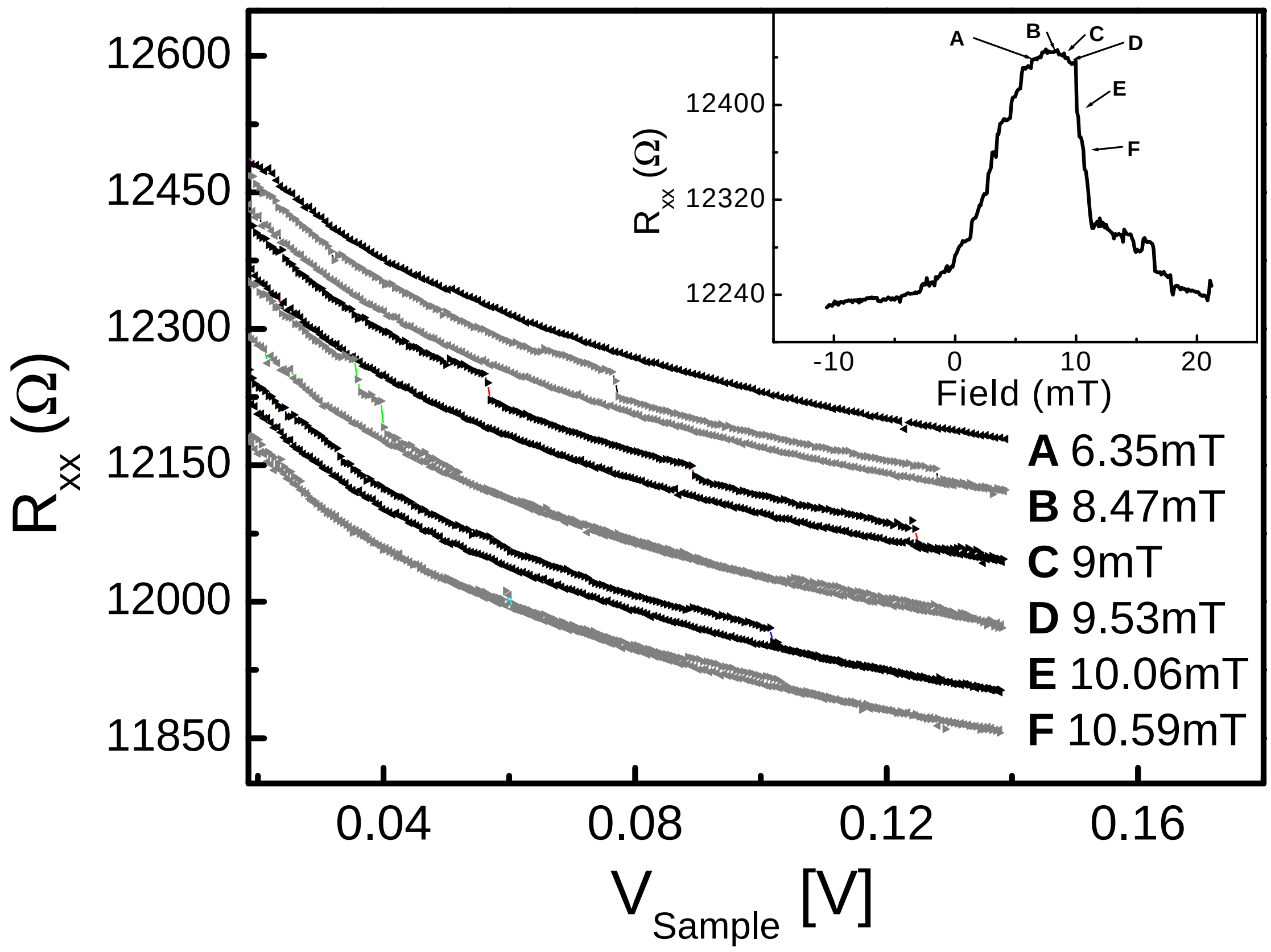

**Fig. 4**